\documentstyle[12pt]{article}                                               
\newcommand{\simlt}  {\raisebox{-.6ex}{$\stackrel{\textstyle <}{\sim}$}}
\newcommand{\simgt}  {\raisebox{-.6ex}{$\stackrel{\textstyle >}{\sim}$}}
\begin{document}                                                                
\begin{flushright}
RAL-TR-97-009 \\                                                         
31 January 1997 \\                                                               
\end{flushright}                                                               
\vspace{0 mm}                                                                   
\begin{center}
{\Large  
Further Evidence for Threefold
Maximal Lepton Mixing and a 
Hierarchical Spectrum of Neutrino
Mass-Squared Differences}            
\end{center}
\vspace{1mm}
\begin{center}                      
{P. F. Harrison\\                   
Physics Department, Queen Mary and Westfield College\\ 
Mile End Rd. London E1 4NS. UK \footnotemark[1]}     
\end{center}                        
\begin{center}
{and}                               
\end{center}                        
\begin{center}                      
{D. H. Perkins\\                      
Nuclear Physics Laboratory, University of Oxford\\
Keble Road, Oxford OX1 3RH. UK \footnotemark[2]} 
\end{center}
\begin{center}                      
{and}                               
\end{center}
\begin{center}                      
{W. G. Scott\\                      
Rutherford Appleton Laboratory\\    
Chilton, Didcot, Oxon OX11 0QX. UK \footnotemark[3]} 
\end{center}
\vspace{1mm}
\begin{abstract}
\baselineskip 0.6cm
The phenomenological case for threefold maximal lepton mixing
is strengthened by more detailed comparison with existing data.
The atmospheric neutrino data are self-consistent and in good
agreement with the threefold maximal mixing prediction $R = 2/3$
($\chi^2$/DOF $=4.3/5$, CL $=51$\%).
Including partially-contained event data, the zenith angle
dependence also points to threefold maximal mixing, 
ruling out a range of mixing schemes, 
including that of Fritzsch and Xing.
Accounting for {\it de facto} uncertainties 
in solar neutrino fluxes,
the solar neutrino data (including HOMESTAKE) are 
shown to be consistent with threefold maximal mixing
($\chi^2$/DOF $= 5.1/3$, CL $=16$\%), 
as is the upper limit from the LSND appearance experiment.
Detailed predictions for forthcoming long-base-line reactor
and accelerator experiments are presented.

\end{abstract}
\begin{center}
{\em To be published in Physics Letters B}
\end{center}
\footnotetext[1]{E-mail:p.f.harrison@qmw.ac.uk}
\footnotetext[2]{E-mail:d.perkins1@physics.oxford.ac.uk}
\footnotetext[3]{E-mail:w.g.scott@rl.ac.uk}
\newpage 
\baselineskip 0.6cm

\noindent{1. INTRODUCTION}
\vskip 0.3cm

The notion of maximal lepton mixing \cite{BRUNO}
and threefold maximal lepton mixing in particular \cite{WOLF1}
is currently exciting renewed interest \cite{HPS} \cite{KIM}.
In the case of threefold maximal mixing 
the mixing matrix may be written:
\begin{eqnarray}
     \matrix{    \nu_e \hspace{1.4cm} 
               & \nu_{\mu} \hspace{1.7cm} 
               & \nu_{\tau}  \hspace{1.0cm} } \nonumber \\
\matrix{ \nu_1 \hspace{0.2cm} \cr 
         \nu_2 \hspace{0.2cm} \cr 
         \nu_3 \hspace{0.2cm} }
\left( \matrix{ \omega / \sqrt{3} & 
                      \omega^2 / \sqrt{3} & 
                              1 / \sqrt{3} \cr
                \omega^2 / \sqrt{3} & 
                     \omega / \sqrt{3} & 
                            1 / \sqrt{3} \cr
      \hspace{5mm} 1/\sqrt{3} \hspace{5mm} & 
         \hspace{5mm}  1/\sqrt{3} \hspace{5mm} & 
           \hspace{5mm} 1/\sqrt{3} \hspace{5mm} \cr } \right)
\end{eqnarray}
\newline 
\noindent
with $\nu_1$, $\nu_2$ and $\nu_3$
the neutrino mass eigenstates,
$\nu_e$, $\nu_{\mu}$ and $\nu_{\tau}$
the neutrino flavour eigenstates,
and $\omega$ a complex 
cube-root of unity.

In Ref.\ \cite{HPS} 
we outlined the phenomenology
of neutrino oscillations 
in a threefold maximal mixing scenario
with a hierarchical neutrino mass spectrum, 
which may explain the atmospheric 
\cite{KAMA} \cite{IMB} \cite{FREJUS} 
\cite{NUSEX} \cite{SOUDAN} and solar
\cite{KAMS} \cite{HOME} \cite{SAGE} 
\cite{GALLEX} neutrino deficits
(the scenario discussed by Giunti et al.\ \cite{KIM} 
is closely related to ours but differs in detail 
regarding the light neutrino masses).
The key feature is that,
independent of the initial neutrino flavour $l$ ($l = e, \mu, \tau$)
the flavour survival probability $P(l \rightarrow l)$,
averaged over neutrino energies, is equal to $5/9$,
over an extremely wide range of $L/E$ values
($L$ is the propagation length and $E$ is the neutrino energy).
This range extends from $L/E \simeq 10^2$ m/MeV,
based on the terrestrial data 
(mainly the atmospheric results,
but including the reactor and accelerator data)
to $L/E$ $\simgt$ $10^{12}$ m/MeV, 
based on the solar data.
The corresponding appearance probability $P(l \rightarrow l')$
to produce a neutrino of a different flavour $l'$ ($l' \ne l$) 
is equal to $2/9$ independent of the initial and final
neutrino flavours, over the same $L/E$ range.
The particular numerical ratios $5/9$ and $2/9$
reflect the democracy of couplings
intrinsic to threefold maximal mixing (Eq.\ 1).
The wide range in $L/E$ over which
the measured probabilities
remain at these values is the experimental signature 
of a hierarchical spectrum of
mass-squared differences for the neutrinos.

Of course, at sufficiently small $L/E$ values
$P(l \rightarrow l) =1$, consistent with ordinary
lepton flavour conservation.
As detailed in Ref.\ \cite{HPS} an abrupt `threshold'
in $L/E$ marks the transition from $P(l \rightarrow l)=1$
to $P(l \rightarrow l) = 5/9$.
It is the location of this threshold on the $L/E$ scale
($L/E \simeq 10^2$ m/MeV) which 
fixes the largest neutrino mass-squared difference
$\Delta m^2 \simeq (0.72 \pm 0.18) \times 10^{-2}$ eV$^2$.
If none of the three neutrino mass eigenstates 
are degenerate then $P(l \rightarrow l) = 1/3$ asymptotically,
and there will be a second threshold
at some larger $L/E$ value, marking the descent from
$P(l \rightarrow l) = 5/9$ to $P(l \rightarrow l) = 1/3$.
The second threshold (if it exists)
fixes the smaller neutrino mass-squared 
difference $\Delta m'^2$.
We found no convincing evidence
for a second threshold over the $L/E$ range 
up to and including the solar data and placed the limit 
$\Delta m'^2 < 0.90 \times 10^{-11}$ eV$^2$
at 90\% confidence (based essentially on the data 
from the two gallium experiments \cite{SAGE} \cite{GALLEX}). 
The above results establish
that the spectrum of mass-squared differences 
for the neutrinos is hierarchical ($\Delta m'^2 \ll \Delta m^2$).
On the assumption that the neutrino masses themselves
exhibit a hierarchical spectrum
(like that of the charged leptons and quarks)
they may be re-expressed in terms
of the neutrino masses ($m_1 < m_2 < m_3$) as follows:
$m_3 \simeq 85 \pm 10$ meV, 
$m_1,m_2 < 3$ $\mu$eV at 90\% confidence 
(see Ref.\ \cite{HPS} for details of fits).

On the basis of the above mass spectrum,
the neutrinos are
too light to be considered useful 
hot dark matter candidates \cite{HDM}.
The mass of the heavy neutrino $\nu_3$
is consistent with a `see-saw' formula \cite{GELL}
relating the masses of the neutrinos 
to the masses of the up-type quarks,
for a heavy (right-handed) majorana mass 
$M_R \simeq 3.6 \times 10^{14}$ GeV
(a see-saw relation involving the charged lepton masses
would yield a smaller value of $M_R$, by a factor of $10^4$).
The above results for the neutrino masses
may also be quoted in terms of the corresponding
compton wavelengths, viz.\ $\lambda_3 \simeq 15$ $\mu$m,
$\lambda_1$, $\lambda_2 > 40$ cm.

The apparent degeneracy or near-degeneracy
of the two light neutrinos $\nu_1$ and $\nu_2$ makes
further progress very difficult.
As long as $\nu_1$ and $\nu_2$
remain effectively degenerate,
only the couplings of the heavy neutrino $\nu_3$
are probed and not those of $\nu_1$ and $\nu_2$ separately.
Hitherto unproposed {\it ad hoc} 
mixing schemes in which the $\nu_3$ is maximally-mixed
(ie.\ has $\nu_e$, $\nu_{\mu}$, $\nu_{\tau}$ content
$1/3$, $1/3$, $1/3$) but which are otherwise arbitrary 
have identical phenomenology to threefold maximal mixing
over the $L/E$ range explored 
in the atmospheric and solar neutrino experiments.
The complete set of such schemes is readily generated 
starting from the threefold maximal mixing matrix (Eq.\ 1)
by forming linear combinations of the first two rows,
with arbitrary complex coefficients
($c_{\theta}e^{i\phi}$, $-s_{\theta}e^{-i\phi}$) and
($s_{\theta}e^{i\phi}$, $c_{\theta}e^{-i\phi}$) 
for $\nu_1$ and $\nu_2$ respectively:
\begin{eqnarray}
     \matrix{    \nu_e \hspace{4.0cm} 
               & \nu_{\mu} \hspace{3.3cm} 
               & \nu_{\tau}  \hspace{1.9cm} } \nonumber \\     
\matrix{ \nu_1 \hspace{0.2cm} \cr 
         \nu_2 \hspace{0.2cm} \cr 
         \nu_3 \hspace{0.2cm} }
\left( \matrix{ 
 (c_{\theta} e^{i\phi} \omega 
     - s_{\theta} e^{-i\phi} \omega^2)/\sqrt{3}   & 
    ( c_{\theta} e^{i \phi} \omega^2
           - s_{\theta} e^{-i \phi} \omega)/\sqrt{3} & 
      ( c_{\theta} e^{i \phi}
           - s_{\theta} e^{-i \phi} )/\sqrt{3} \cr 
 (s_{\theta} e^{i \phi} \omega^2 
        + c_{\theta} e^{-i \phi} \omega)/\sqrt{3} & 
   (s_{\theta} e^{i \phi} \omega
            + c_{\theta} e^{-i \phi} \omega^2)/\sqrt{3} & 
        (s_{\theta} e^{i \phi}
              + c_{\theta} e^{-i \phi} )/\sqrt{3} \cr 
             1/\sqrt{3} & 1/\sqrt{3} & 1/\sqrt{3} \cr } \right)
\end{eqnarray}
\newline 
\noindent
where $c_{\theta} =\cos \theta$ and $s_{\theta} = \sin \theta$.
Threefold maximal mixing predicts $\theta =0$ (mod $\pi/2$)
with $\phi$ a redundant phase.
More generally
$\theta$ and $\phi$ are both measurable.
Data on neutrinos from supernovae
($L/E$ $\simgt$ $10^{20}$ m/MeV)
although presently too sparse to yield 
reliable information on mixing \cite{KRAUSS}
might eventually measure the couplings
of the light neutrinos and test the above prediction. 
If $\nu_1$ and $\nu_2$ are truly degenerate
(as is not excluded by the present data)
then all such mixing schemes are equivalent
and the mixing may then be defined
to be threefold maximal by choice.

On the other hand, 
proposed mixing schemes in which one (and only one) 
of the neutrino {\em flavour} eigenstates is maximally-mixed
(ie.\ has $\nu_1$, $\nu_2$, $\nu_3$ content $1/3$, $1/3$, $1/3$)
can in general be distinguished by the data 
in the $L/E$ range currently accessed
(so that it matters whether the postulated maximally-mixed
state constitutes a `row' or `column' of the mixing-matrix).
The complete set of real mixing-matrices 
with the $\nu_e$ maximally-mixed 
has been discussed by Acker et al.\ \cite{ACKER},
while a particular (real) mixing-matrix with the $\nu_{\tau}$
maximally-mixed is singled out in a recent paper
by Fritzsch and Xing \cite{FRITZSCH}.
The complete set of mixing matrices with the
$\nu_{\tau}$ maximally mixed for example
(as in the Fritzsch-Xing ansatz)
is readily generated analogously to Eq.\ 2: 
\begin{eqnarray}
     \matrix{    \nu_e \hspace{3.9cm} 
               & \nu_{\mu} \hspace{2.8cm} 
               & \nu_{\tau}  \hspace{1.0cm} } \nonumber \\     
\matrix{ \nu_1 \hspace{0.2cm} \cr 
         \nu_2 \hspace{0.2cm} \cr 
         \nu_3 \hspace{0.2cm} }
\left( \matrix{ 
 (c_{\theta} e^{i\phi} \omega 
     - s_{\theta} e^{-i\phi} \omega^2)/\sqrt{3}   & 
    ( c_{\theta} e^{i \phi} \omega^2
           + s_{\theta} e^{-i \phi} \omega)/\sqrt{3} & 
               \hspace{6mm} 1/\sqrt{3} \hspace{6mm} \cr 
 (c_{\theta} e^{i \phi} \omega^2 
        - s_{\theta} e^{-i \phi} \omega)/\sqrt{3} & 
   (c_{\theta} e^{i \phi} \omega
            + s_{\theta} e^{-i \phi} \omega^2)/\sqrt{3} & 
               \hspace{6mm} 1/\sqrt{3} \hspace{6mm} \cr 
           ( c_{\theta} e^{i \phi}
           - s_{\theta} e^{-i \phi} )/\sqrt{3}   &
         (c_{\theta} e^{i \phi}
              + s_{\theta} e^{-i \phi} )/\sqrt{3} & 
               \hspace{6mm} 1/\sqrt{3} \hspace{6mm} \cr } \right)
\end{eqnarray}
\newline
\noindent
where the Fritzsch-Xing ansatz is reproduced 
in the case $\theta = \pi/4$ with $\phi=0$. 
While such schemes are also 
fairly described as {\it ad hoc}
they do yield valid candidate mixing matrices
which can be tested against experiment 
(see Section~2 and Section~3 below).

It is hardly necessary to emphasise here
that the notion of large mixing 
in the lepton sector
is by no means a universal tenet.
On the contrary, the widely accepted 
MSW solution \cite{WOLF2} \cite{MS} \cite{BETHE}
to the solar neutrino problem currently
favours small mixing angles in the lepton sector,
as do recent results from the LSND \cite{LSND} 
appearance experiment (see Section~3 and Section~4 below).

In this paper 
we re-examine the experimental evidence 
for (and against) threefold maximal lepton mixing,
in comparison with a range of
more general mixing schemes.
Detailed predictions for forthcoming
reactor and accelerator experiments are given,
based on the threefold maximal mixing scenario.

\vskip 0.3cm
\noindent{2. THE ATMOSPHERIC DATA}
\vskip 0.3cm

The best measured quantity in the
atmospheric neutrino experiments is 
the atmospheric neutrino ratio 
$R = (\mu/e)_{DATA}/(\mu/e)_{MC}$,
where $(\mu/e)_{DATA}$ is the measured
ratio of muon to electron events,
and $(\mu/e)_{MC}$ is the expected
ratio of muon to electron events,
assuming no oscillations.
Results for $R$ 
for fully contained events
from the various underground experiments
are summarised in Table~1.
The KAMIOKA and IMB experiments are
based on the water-Cerenkov technique,
while FREJUS, NUSEX and SOUDAN
are ionisation-based tracking detectors. 
A common $\pm 5$\% 
uncertainty in the predicted flux ratio 
has been subtracted from the quoted
systematic errors where appropriate,
so that the errors given here should
be largely independent
experiment-to-experiment.
The same data are 
plotted in Figure~1, where
the errors shown combine the 
above statistical and systematic 
errors in quadrature.

Contrary to what has been asserted 
(eg.\ Ref.\ \cite{WINTER}),
there is no evidence for a discrepancy 
between the results of the water-Cerenkov 
and tracking detectors, 
at least for the contained event data.
A weighted mean over the five experiments 
gives $R = 0.64 \pm 0.06$
($\chi^2$/DOF$=4.2/4$, CL $=38$\%).
In the threefold maximal mixing scenario
one expects $R = 2/3$ \cite{HPS} for contained events,
in the approximation that the flux ratio at production
$F = \phi(\nu_{\mu}+\bar{\nu}_{\mu})/\phi(\nu_{e}+\bar{\nu}_{e})$ 
($\equiv \nu_{\mu}/\nu_e$) is equal to $2/1$ at low energy.
In Figure~1 the broken line shows the
threefold maximal mixing prediction $R=2/3$.
The contained event data are clearly consistent
with this ($\chi^2$/DOF $=4.3/5$, CL $=51$\%),
the higher confidence level for the maximal mixing hypothesis
reflecting the reduction in the number of fitted parameters.
Furthermore, in the threefold maximal mixing scenario,
the final $\nu_e$ rate turns out to be unaffected 
by the oscillations (in the approximation $\nu_{\mu}/\nu_e=2/1$),
so that it is the $\nu_{\mu}$ rate 
which is expected to be reduced by a factor of $2/3$,
which is also in agreement with observation \cite{DHP}
for the $\nu_{\mu}$ and $\nu_e$ rates separately. 

The KAMIOKA partially-contained sample
(the so-called `multi-GeV' data-set)
provides a measurement of the zenith angle
dependence of $R$ and hence the $L/E$ dependence.
Although there is no evidence for neutrino oscillations 
in the atmospheric data for $L/E < 10^2$ m/MeV,
the data with $L/E > 10^2$ m/MeV
can be utilised to yield independent information on
survival {\em and} appearance probabilities,
exploiting the dependence of the initial 
$\nu_{\mu}/\nu_e$ flux ratio
on energy and zenith angle.
Figure~2a shows the measured value 
of the atmospheric neutrino ratio $R$ 
plotted versus the initial flux ratio $F$ 
for all the data with $L/E > 10^2$ m/MeV.
The point plotted at $F = 2.2$
is the overall average of the contained event data, 
discussed above.
The points at larger values of $F$ 
are based on the KAMIOKA partially-contained 
(multi-GeV) data-set.
The value of $F$ for each point was calculated 
(as a function of energy and zenith angle) 
using a Monte Carlo simulation
of the propagation and decay 
of pions and muons in the atmosphere.

In the presence of neutrino oscillations
the predicted value of $R$ as a function of $F$ is given by:
$R=    (P(e \rightarrow \mu)+F \times P(\mu \rightarrow \mu))/
       (F \times P(e \rightarrow e)+F^2 \times P(\mu \rightarrow e))$.
In the maximal mixing scenario
$P(\mu \rightarrow \mu) = P(e \rightarrow e) = 5/9$ and
$P(e \rightarrow \mu) = P(\mu \rightarrow e) = 2/9$
leading to the dependence shown by the dashed curve in Figure~2a,
which is clearly consistent with the data
($\chi^2$/DOF $= 1.2/2$, CL $=55$\%).
In the Fritzsch-Xing model (see Section~1) \cite{FRITZSCH},
$P(\mu \rightarrow \mu) = 5/9$ but $P(e \rightarrow e) =1$ 
and $P(\mu \rightarrow e)=P(e \rightarrow \mu)=0$
for the atmospheric data,
so that $R=5/9$ independent of $F$ as shown by the dotted line.
The data rule out the Fritzsch-Xing model 
with better than 99\% confidence
($\chi^2$/DOF $=9.4/2$, CL $= 0.7$\%).

The full range of mixing schemes with
the $\nu_{\tau}$ maximally mixed (Eq.\ 3)
may also be tested against experiment
using the data of Figure~2a. 
All these schemes are as well motivated
a priori as the Fritzsch-Xing model itself.
It turns out that 
in the $L/E$ range between the two thresholds, ie.\ for 
$(\Delta m^2/2)^{-1}$ $\simlt$ $L/E$ $\simlt$ $(\Delta m'^2/2)^{-1}$,
observables depend only on the product $\sin 2 \theta \cos 2 \phi$.
Specifically
$P(\mu \rightarrow \mu) = 1/2+2/9(1/2 - \sin 2 \theta \cos 2 \phi)^2$,
$P(e \rightarrow e) = 1/2+2/9(1/2 + \sin 2 \theta \cos 2 \phi)^2$ and
$P(e \rightarrow \mu) = P(\mu \rightarrow e) 
= 2/9(1-\sin^2 2 \theta \cos^2 2 \phi)$.
Beyond the second threshold ($L/E$ $\simgt$ $(\Delta m'^2/2)^{-1}$)
if present,
$P(\mu \rightarrow \mu) = P(e \rightarrow e) = 1/3(1+1/2\sin^2 2 \theta)$ 
and
$P(\mu \rightarrow e) = P(e \rightarrow \mu) = 1/3(1-1/2\sin^2 2 \theta)$.
Remaining probabilities (involving the $\tau$) 
follow immediately in each case, 
since the rows and columns
of the probability matrices sum to unity.

On the assumption
that the neutrino mass spectrum is hierarchical
($\Delta m'^2 \ll \Delta m^2$)
the formulae valid between the two thresholds apply.
The calculated $\chi^2$ is plotted as a function
of $\sin 2 \theta \cos 2 \phi$ in Figure~2b.
Threefold maximal mixing 
predicts $\sin 2 \theta \cos 2 \phi = 0$,
while the Fritzsch-Xing model
corresponds to $\sin 2 \theta \cos 2 \phi = 1$.
The best fit yields 
$\sin 2 \theta \cos 2 \phi = 0.07 \pm$
\raisebox{0.9ex}{0.28} \raisebox{-0.9ex}{\hspace{-8.9mm}0.15}.
Thus most of these schemes 
are ruled out by the data, 
along with the Fritzsch-Xing model,
while threefold maximal mixing
is both fully consistent with the data 
and very close to the preferred solution.

\vskip 0.3cm
\noindent{3. THE SOLAR DATA}
\vskip 0.3cm

Any analysis of the existing solar data
requiring a knowledge of the solar neutrino fluxes,
relies on predictions from solar models.
In Figure~3, the Standard Solar Model predictions of
Proffitt (PR) \cite{PR}, 
Bahcall and Pinsonneault (BP) \cite{BP},
Turk-Chiese and Lopes (TL) \cite{TCL}, 
and Dar and Shaviv (DS) \cite{DS}
are compared side-by-side,
by plotting the ratio of observed to predicted rates
in each of the four solar neutrino experiments,
KAMIOKA \cite{KAMS}, HOMESTAKE \cite{HOME},
SAGE \cite{SAGE} and GALLEX \cite{GALLEX}.
The KAMIOKA and HOMESTAKE experiments
are primarily sensitive to $^8$B neutrinos,
and the resulting suppression factors are seen to be
strongly dependent on the flux calculation used.
On the other hand, in the two gallium experiments, 
SAGE \cite{SAGE} and GALLEX \cite{GALLEX},
the main contribution to the rate is from pp neutrinos, 
and the corresponding dependence is seen to be small. 
Furthermore, the efficiency of both the 
SAGE and GALLEX detectors has been measured
(to $\sim 10$\% accuracy)
using a laboratory neutrino source ($^{51}$Cr).

In view of the particular stability 
and reliability of the gallium data,
a number of conclusions follow immediately. 
In Figure~3 the dashed line 
$P(e \rightarrow e) = 5/9$
is the threefold maximal mixing prediction
applying in the case that the second threshold 
is taken to lie beyond the solar data in $L/E$,
as is assumed in Ref.\ \cite{HPS}.
It is clear that the results
from the two gallium experiments 
are in excellent agreement with this.
The mean of the SAGE and GALLEX results
averaged over the four flux models
yields $P(e \rightarrow e) = 0.55 \pm 0.05$,
to be compared with the $5/9$ expected.
The scenario inferred by Giunti et al.\ \cite{KIM}
with the second threshold
coinciding in $L/E$ with the solar data
(leading to $\Delta m'^2 \simeq 10^{-10}$ eV$^2$ \cite{KIM})
does not now seem well-founded, and 
in the same way the scenario described
by Acker et al.\ \cite{ACKER} (see Section~1) 
which places all the solar data beyond the second threshold
($\Delta m'^2 \simeq \Delta m^2 \simeq 10^{-2}$ eV$^2$ \cite{ACKER})
is also now excluded by the gallium data
(both Giunti et al.\ and Acker et al.\ would predict 
$P(e \rightarrow e) \simeq 1/3$ in the gallium experiments).

It should be added 
that we have not analysed the
complete set of models with the $\nu_e$ maximally mixed,
as in the model of Acker et al.
The relevant formulae follow immediately 
from those given for $\nu_{\tau}$ maximal mixing, 
under the interchange $e \leftrightarrow \tau$.
It may be remarked that putting $\theta = \pi/4$
then yields the set of real mixing matrices
discussed by Acker et al.\ and that
putting $\phi = \pi/12$ (in addition) gives a real matrix
with the $\nu_3$ (as well as the $\nu_e$) maximally mixed.
This is {\em one} of the infinite set of matrices (Eq. 2) 
which reproduce the phenomenology
of threefold maximal mixing up to 
the second threshold which were discussed in Section~1. 
We do not consider this particular example
to be of any special interest.

Coming now to the KAMIOKA and HOMESTAKE results,
we previously concluded \cite{HPS} 
that the KAMIOKA point was consistent with  
$P(e \rightarrow e) = 5/9$ and 
that the HOMESTAKE point was low,
relying on an average of the BP and TL flux calculations.
From Figure~3 it is apparent that,
if one uses the DS fluxes,
it is the HOMESTAKE point which is consistent with $5/9$,
and the KAMIOKA point which is high.
That such large differences should exist
between the different flux calculations,
is clearly very unsatisfactory.
The predicted $^8$B flux varies by more than
a factor of two between the most extreme models,
which is far outside the error limits quoted 
by the authors of the individual calculations,
with no indication of an emerging consensus
in the literature.
Under the circumstances, 
it would seem prudent \cite{PETCOV}
to avoid drawing conclusions 
which depend on the predicted $^8$B flux, 
if at all possible.

New results are presented here, 
obtained by refitting the existing solar data 
taking the $^8$B flux as an adjustable parameter.
In so far as they are known,
correlations between the $^8$B flux
and the other flux components are taken into account
in these fits by linearly interpolating
the other flux components,
between the above four flux calculations,
as a function of the $^8$B flux.
In addition to the threefold maximal mixing solution,
all of the well-known
$2 \times 2$ solutions to the solar neutrino problem,
viz.\ small-angle MSW, large-angle MSW and the
`just-so' vacuum oscillation solution \cite{GLASH},
are found to survive as distinct $\chi^2$ minima in these fits
(matter effects, essentially inoperative 
in the case of maximal mixing \cite{HPS},
are incorporated where needed here
using standard $2 \times 2$ formulae \cite{KP1}).
Best-fit parameters and confidence levels 
are of course modified, 
with respect to those usually quoted, 
in most cases.

The results of these fits are plotted in Figure~4,
as a function of $L/E$ \cite{HPS}.
Figure~4a shows the threefold maximal mixing fit,
which assumes $P(e \rightarrow e) =5/9$ independent of energy.
Further details of this fit are given in Table~2.
The best-fit value for the total $^8$B flux
is $\Phi(^8$B) $= 3.4 \times 10^6$ cm$^2$ s$^{-1}$ 
($\chi^2$/DOF $= 5.1/3$, CL $= 16$\%),
which lies between the TL and DS predictions.
Figure~4b shows the small angle MSW fit,
which gives $\Phi(^8$B) $= 4.4 \times 10^6$ cm$^2$ s$^{-1}$
(essentially the TL prediction), with
$\sin \theta = 0.030$, $\Delta m^2 = 0.61 \times 10^{-5}$ eV$^2$
($\chi^2$/DOF $= 0.004/1$, CL $= 96$\%).
For the large angle MSW
and `just-so' vacuum oscillation solutions,
the best-fit $^8$B flux turns out to be larger/smaller
than the most extreme model in each case,
viz.\ the PR and DS models respectively.
To avoid relying on extrapolated fluxes,
the results for the large angle MSW fit (Figure~4c)
are given using the PR flux, viz.\  
$\Phi(^8$B) $= 6.5 \times 10^6$ cm$^2$ s$^{-1}$,
$\sin \theta = 0.50$, $\Delta m^2 = 2.1 \times 10^{-5}$ eV$^2$
($\chi^2$/DOF $= 1.8/2$, CL $= 41$\%).
Similarly for the `just-so' vacuum oscillation fit (Figure 4d)
the DS flux is used, viz.\
$\Phi(^8$B) $= 2.8 \times 10^6$ cm$^2$ s$^{-1}$,
$\sin \theta = 0.71$, $\Delta m^2 = 1.8 \times 10^{-10}$ eV$^2$ 
($\chi^2$/DOF $= 2.2/2$, CL $= 34$\%).
In each of Figures 4a-d 
the data points are plotted 
using the above flux values,
with a correction \cite{HPS} applied to the KAMIOKA point
to account for the neutral-current contribution.
The solid curve in each case is the expected 
suppression as a function of $L/E$,
corresponding to the fitted mixing parameters.

As is well-known, the MSW effect
is observable only over a limited range
in parameter space, corresponding to 
about four orders of magnitude in
neutrino mass-squared difference 
(or two orders of magnitude
in terms of neutrino mass).
In Figures~4b-4d
the broken curves show the effect of rescaling 
the relevant neutrino mass by a factor 100 (up or down) 
with respect to the best-fit value, 
keeping the mixing angle fixed.
The expectations are scaled,
or translated on the logarithmic $L/E$-scale,
proportional (to the square of) the mass rescaling.
The range of observability of the MSW effect
is determined by the maximum width of the `bathtub'
(attained in the case of the large angle solution).
Similarly for the vacuum oscillation solution, 
with the solar data spanning
less than two orders of magnitude on the $L/E$ scale,
the associated threshold is
observable only if the neutrino 
mass-squared difference is `just-so'.
While an element of `fine-tuning' undoubtedly enters therefore,
for both the MSW and vacuum oscillation solutions
(which is not present in the case of threefold maximal mixing), 
it is difficult to know how to take
this into account quantitatively.
The conclusions which follow are based simply
on the straightforward confidence levels quoted above.

Based on the above confidence levels,
the small angle MSW fit continues 
to give the best account of the solar data,
with both the large angle MSW
and `just-so' vacuum oscillation solutions
also having higher fit-probabilities than
threefold maximal mixing, 
albeit by somewhat smaller factors.
On the other hand,
threefold maximal mixing is {\em not} excluded
by the solar data (including the HOMESTAKE point) 
even at 90\% confidence.
A clear prediction of the threefold
maximal mixing scenario, 
independent of the predicted $^8$B flux, 
is that the KAMIOKA and HOMESTAKE experiments
must ultimately measure the same suppression.
Presumably future solar experiments,
in particular SNO \cite{SNO},
will eventually settle the question of the $^8$B flux.
Meanwhile, data from the new KAMIOKA detector
(SUPER-K \cite{SUPERK})
will obviously be of very considerable interest.
 
\vskip 0.3cm
\noindent{4. ACCELERATOR AND REACTOR EXPERIMENTS}
\vskip 0.3cm

The LSND \cite{LSND} collaboration claim a positive signal
for $\bar{\nu}_{\mu} \rightarrow \bar{\nu}_e$
in an appearance experiment ($L/E \simeq 0.98$ m/MeV),
with an appearance probability 
$P(\bar{\mu} \rightarrow \bar{e}) \simeq (0.31 \pm 0.10 \pm 0.05)$\%.
The LSND result is plotted in Figure~5,
together with the upper limits given by
KARMEN \cite{KARMEN} ($L/E \simeq 0.44$ m/MeV) 
and BNL-776 \cite{BNL776} ($L/E \simeq 0.71$ km/GeV).
The curve shows the predicted appearance 
probability in threefold maximal mixing, 
plotted as a function of $L/E$, assuming
$\Delta m^2 \simeq 0.72 \times 10^{-2}$ eV$^2$ \cite{HPS}.
Taken at face value, the LSND result 
would immediately exclude threefold maximal mixing
(at least with $\Delta m^2 \simeq 0.72 \times 10^{-2}$ eV$^2$).
Of course, the LSND result can be accommodated
together with the atmospheric and solar results
in a more general mixing scheme \cite{ACK2}.
It should be pointed out, however, that
the LSND result is only marginally consistent
with the upper limits from KARMEN and BNL-776,
and that, furthermore, due to unexplained anomalies
in the spatial distribution of the events in the detector,
the LSND result remains controversial, even within
the LSND collaboration. The upper limit obtained 
by Hill \cite{HILL} from a subset of the LSND data 
is also plotted in Figure~5 and is clearly 
fully consistent with threefold maximal mixing.

In Figure~5, the expected sensitivities 
of the two CERN appearance experiments, 
CHORUS \cite{CHORUS} and NOMAD \cite{NOMAD}
($L/E \simeq 0.05$ km/GeV) are also indicated
(90\% confidence upper limits are shown, 
based on the projected running).
In the threefold maximal mixing scenario
the predicted appearance rates at these $L/E$ values
are extremely small ($P(\mu \rightarrow \tau) \simeq 10^{-7}$).
If threefold maximal mixing is correct,
no tau-lepton events should be seen
in either of these experiments.
 
Assuming threefold maximal mixing is
correct, spectacular effects are expected 
in long-baseline reactor and accelerator
experiments, planned or in progress.
Results on $\bar{\nu}_e$ survival rates
from the CHOOZ \cite{CHOOZ} reactor experiment 
($L/E \simeq 200$ m/MeV) are expected within months, 
and from the PALO-VERDE \cite{PALOV} reactor experiment
($L/E \simeq 150$ m/MeV) within one year.
The MINOS \cite{MINOS} experiment ($L/E \simeq 50$ km/GeV) 
will utilise a $\nu_{\mu}$ beam produced by the FNAL main-injector
in conjunction with a new detector located in the SOUDAN mine,
and is expected to be operational by the year 2001.
MINOS should be sensitive to $\nu_e$ and $\nu_{\tau}$
appearance as well as $\nu_{\mu}$ disappearance.
Similar long-baseline accelerator experiments, 
planned or proposed, include
KEK/KAMIOKA \cite{KEK} and CERN/GRAN-SASSO \cite{CARLO}.
In Figure~6 the threefold maximal mixing
predictions for the CHOOZ, PALO-VERDE and MINOS
experiments are plotted out in full
(experimental resolution effects are not included
in these plots, but are not expected to dominate). 
With both $L$ and $E$ relatively well determined,
each of these experiments should succeed in mapping out 
essentially one full oscillation cycle,
providing the first detailed and definitive proof
of the existence of neutrino flavour oscillations.

\vskip 0.3cm
\noindent{5. CONCLUSION}
\vskip 0.3cm

Having re-examined the available experimental evidence,
the threefold maximal mixing scenario remains our preferred
solution to the problem of lepton mixing.
It provides a good fit {\it ab initio} 
to $26$ data points from $19$ out of $20$ 
disappearance experiments \cite{HPS}.
As discussed in this paper,
the discrepancy with HOMESTAKE 
may now be reconciled.
Regarding the appearance data,
only the LSND result is in disagreement
with threefold maximal mixing.

While the suppression factors
measured in the atmospheric and solar experiments 
are naturally interpreted
in terms of neutrino oscillations
(consistent with
the threefold maximal mixing scenario)
definitive and detailed
proof of the existence of actual neutrino oscillations
can only come from laboratory experiments
with controllable man-made beams,
and precision detectors.
Forthcoming long-baseline projects
at reactors and accelerators 
should, therefore, be decisive.

In the longer term,
attention will perhaps turn to the second threshold
and the measurement of $\Delta m'^2$, where,
as regards the couplings of the two light neutrinos,
experiments sensitive to neutrinos from distant supernovae 
\cite{SMITH} may well be our only hope of progress. 

\vspace{5mm}
\noindent {\bf Aknowledgement}

\noindent We are indebted to J. N. Bahcall, A. Dar, Y. Declais, 
P. I. Krastev, S. T. Petcov and S. Sarkar
for helpful correspondence relating to the topics discussed in this paper.

\newpage

\noindent {\bf {\large Figure Captions}}

\vspace{10mm}
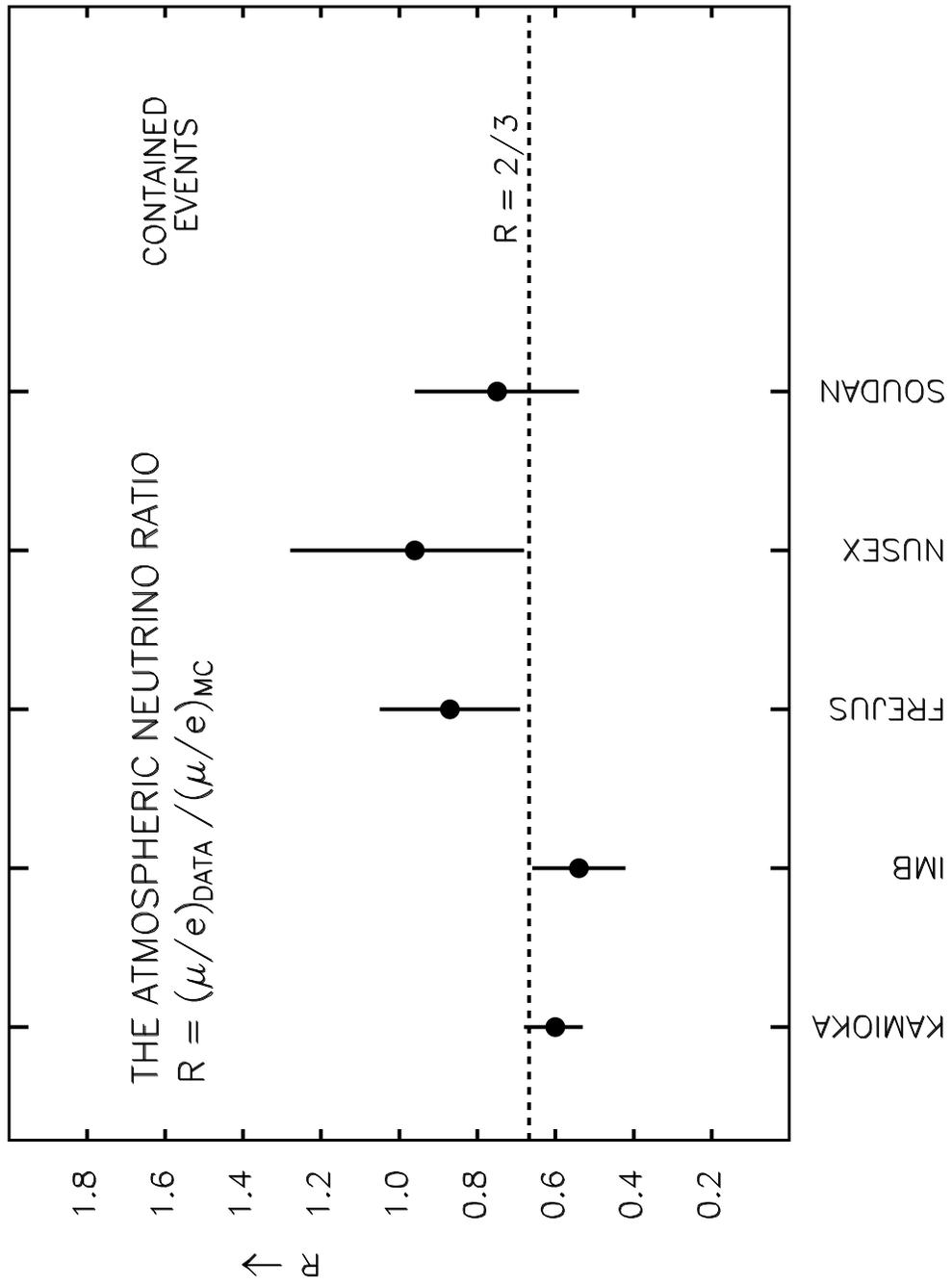
\noindent Figure~1.
The atmospheric neutrino ratio $R = (\mu/e)_{DATA}/(\mu/e)_{MC}$
for contained events from the various underground experiments.
The errors plotted combine in quadrature 
the statistical and systematic errors given in Table~1
and should be largely independent from experiment to experiment.
There is no evidence for a discrepancy between the results
of the water-Cerenkov and the tracking detectors.
The data are consistent with the threefold
maximal mixing prediction $R=2/3$ (broken line).
 
\vspace{10mm}
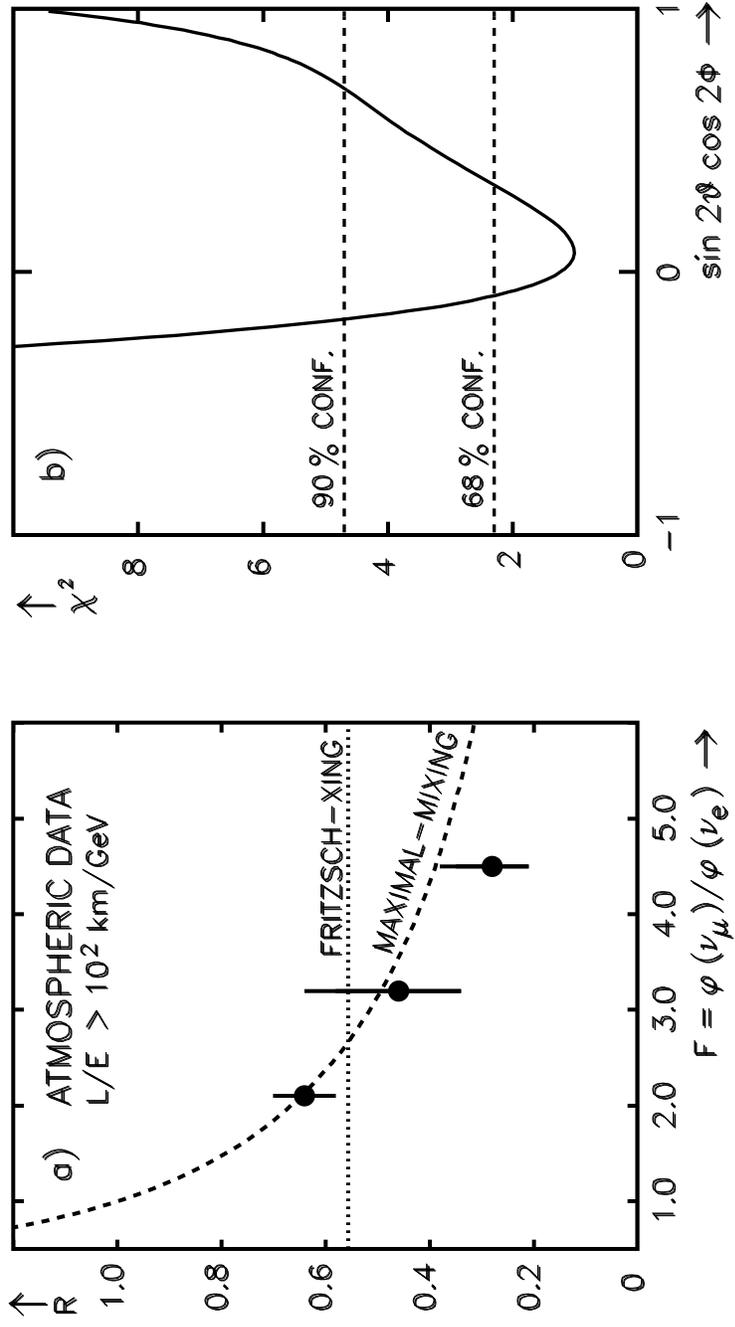
\noindent Figure~2.
a) The atmospheric neutrino ratio $R$ plotted
versus the flux ratio $F=\phi(\nu_{\mu})/$ $\phi(\nu_e)$
for all the data with $L/E > 10^2$ km/GeV.
Threefold maximal mixing predicts the dependence shown
(dashed curve) while the Fritzsch-Xing ansatz 
\cite{FRITZSCH} predicts no dependence (dotted line).
b) The corresponding $\chi^2$ plotted as a function of 
$\sin 2\theta \cos 2\phi$ (see text) for a range of ansatze 
with the tau-lepton maximally-mixed.
The Fritzsch-Xing ansatz corresponds to $\sin 2\theta \cos 2\phi = 1$ 
and is ruled-out by the data, 
while threefold maximal mixing predicts $\sin 2\theta \cos 2\phi = 0$, 
which is close to the $\chi^2$-minimum.
 
\vspace{10mm}
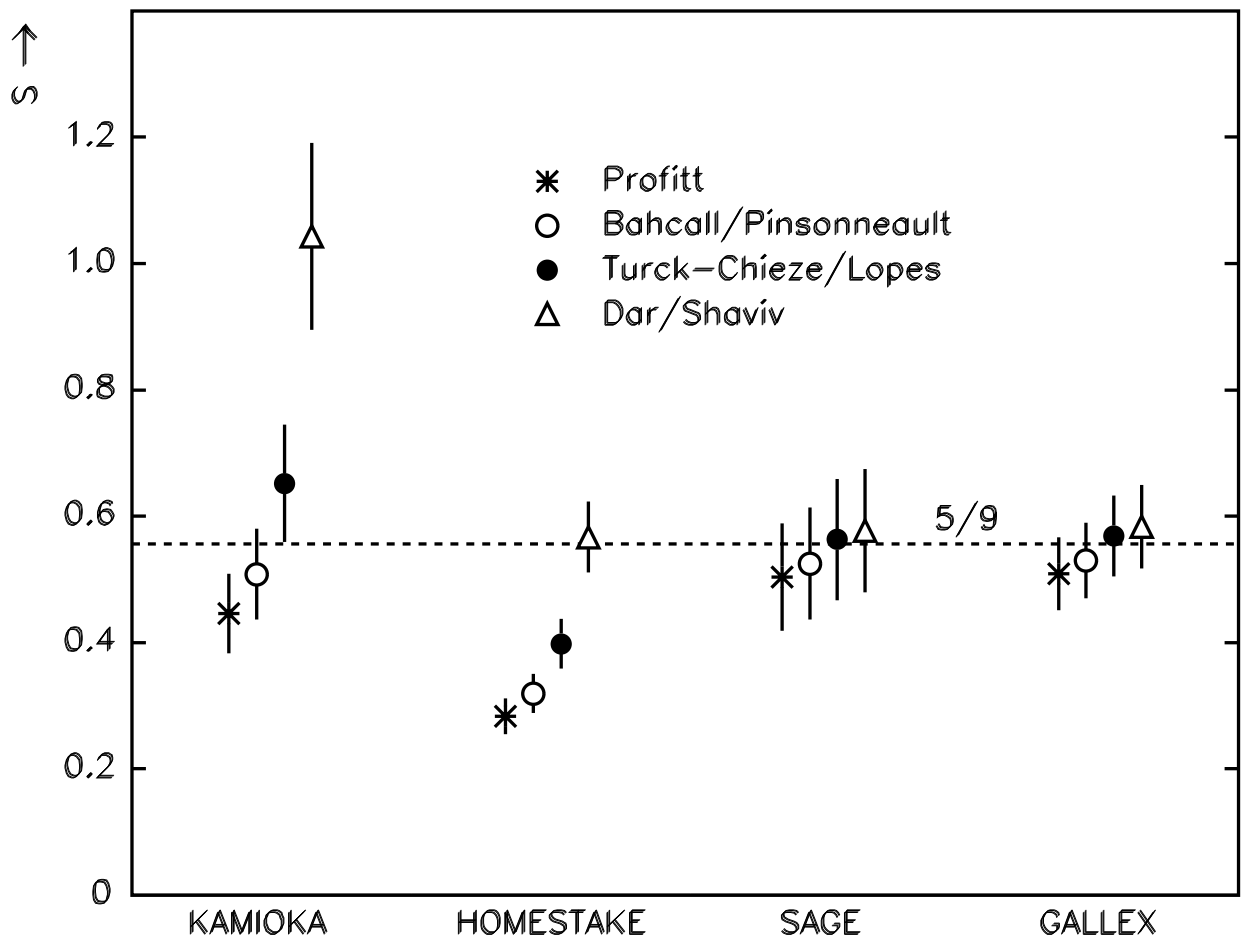
\noindent Figure~3.
The ratio $S$ of observed to predicted rates
in the four existing solar neutrino experiments
for each of four solar model predictions.
The KAMIOKA and HOMESTAKE values depend
principally on the calculated $^8$B flux
and show a wide scatter, while the two gallium
experiments are sensitive mainly to pp neutrinos,
for which the differences between the models are small.
The broken line shows the threefold maximal mixing prediction
$P(e \rightarrow e) = 5/9$ which is in excellent agreement
with the gallium data. 
The scenario of Giunti et al. \cite{KIM} and that of
Acker et al. \cite{ACKER} predict $P(e \rightarrow e) \simeq 1/3$ 
for gallium and are now excluded by the data.

\vspace{10mm}
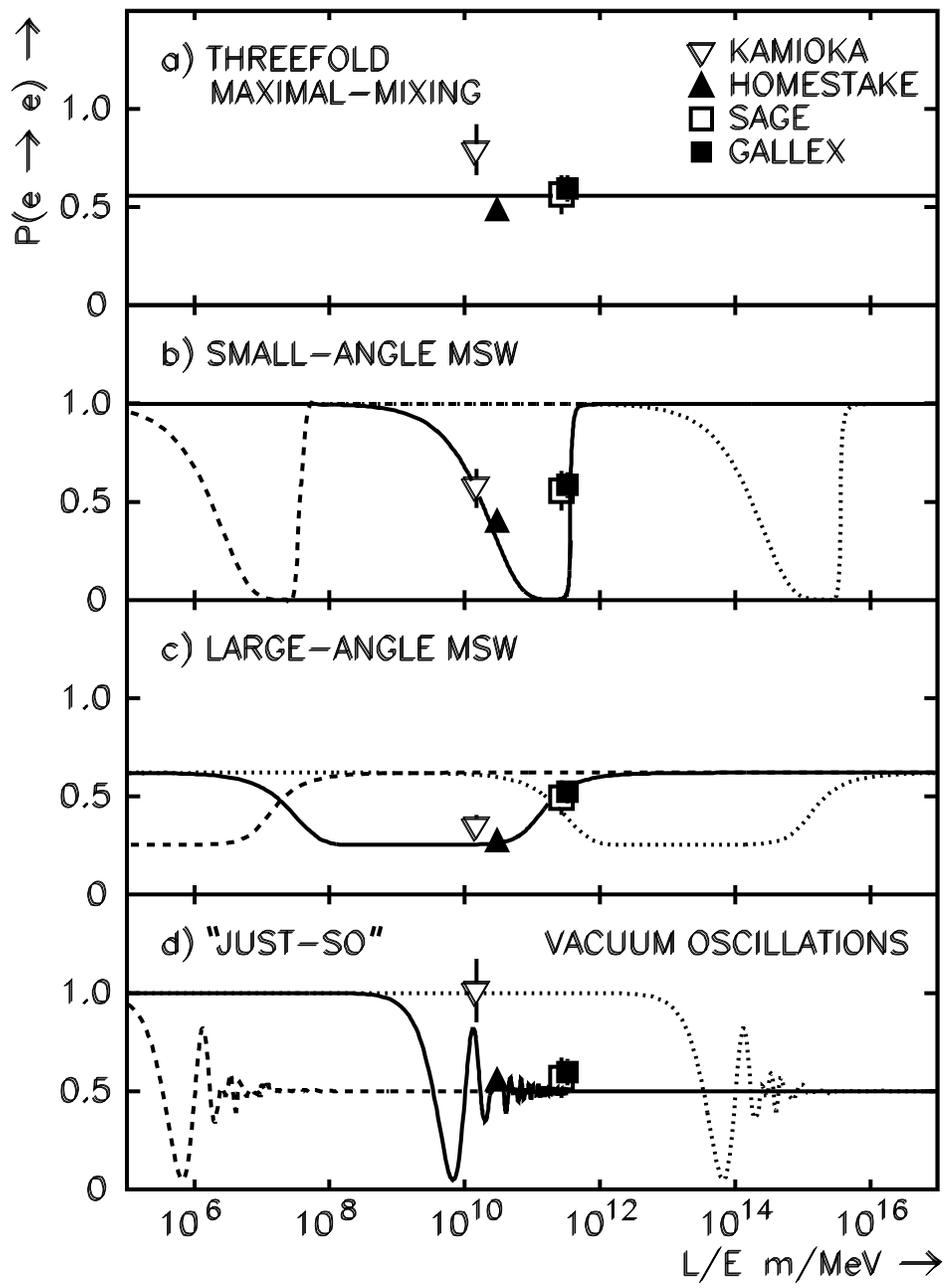
\noindent Figure~4.
Fits to the solar data plotted 
as a function of $L/E$ (solid curves).
The total $^8$B flux has been taken 
as an adjustable parameter in these fits: 
a) $\Phi(^8$B) $= 3.4 \times 10^6$ cm$^2$ s$^{-1}$;
b) $\Phi(^8$B) $= 4.4 \times 10^6$ cm$^2$ s$^{-1}$,
$\sin \theta = 0.030$, $\Delta m^2 = 0.61 \times 10^{-5}$ eV$^2$;
c) $\Phi(^8$B) $= 6.5 \times 10^6$ cm$^2$ s$^{-1}$, 
$\sin \theta = 0.50$, $\Delta m^2 = 2.1 \times 10^{-5}$ eV$^2$
and
d) $\Phi(^8$B) $= 2.77 \times 10^6$ cm$^2$ s$^{-1}$,
$\sin \theta = 0.71$, $\Delta m^2 = 1.8 \times 10^{-10}$ eV$^2$. 
The data points in each case
are calculated using the above flux values (see text).
The broken curves show the effect of varying
the neutrino mass by a factor of 100 up or down
with respect to the best-fit value,
with no dependence in threefold maximal mixing
($P(e \rightarrow e) = 5/9$).

\newpage

\vspace{10mm}
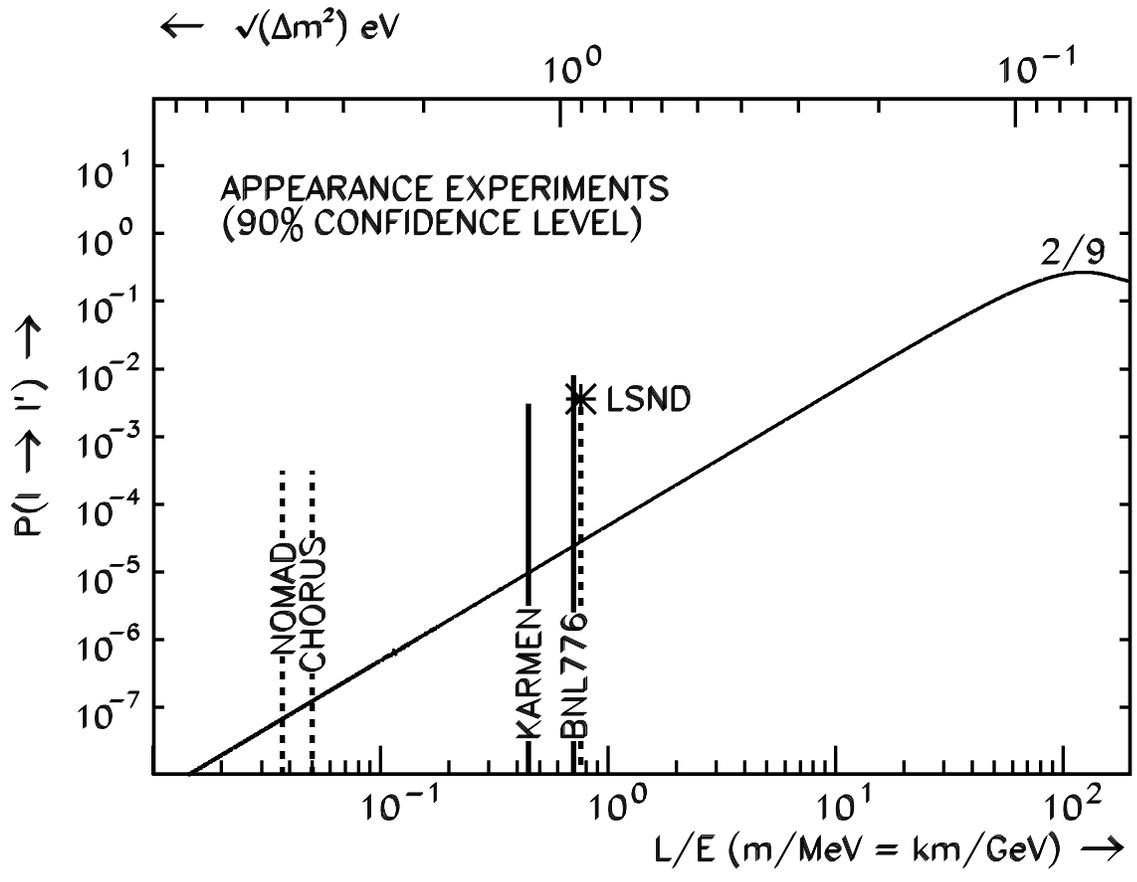
\noindent Figure~5.
The predicted appearance probability $P(l \rightarrow l')$
in the threefold maximal mixing scenario
with $\Delta m^2 = 0.72 \times 10^{-2}$ eV$^2$ \cite{HPS}.
The LSND result \cite{LSND} is indicated together with the
upper limit obtained by Hill \cite{HILL} 
from a subset of the LSND data.
Upper limits from KARMEN \cite{KARMEN} and BNL-776 \cite{BNL776}
are shown, together with the projected sensitivities 
(90\% upper limits, assuming no signal) 
of the CHORUS \cite{CHORUS} and NOMAD \cite{NOMAD} experiments.

\vspace{10mm}
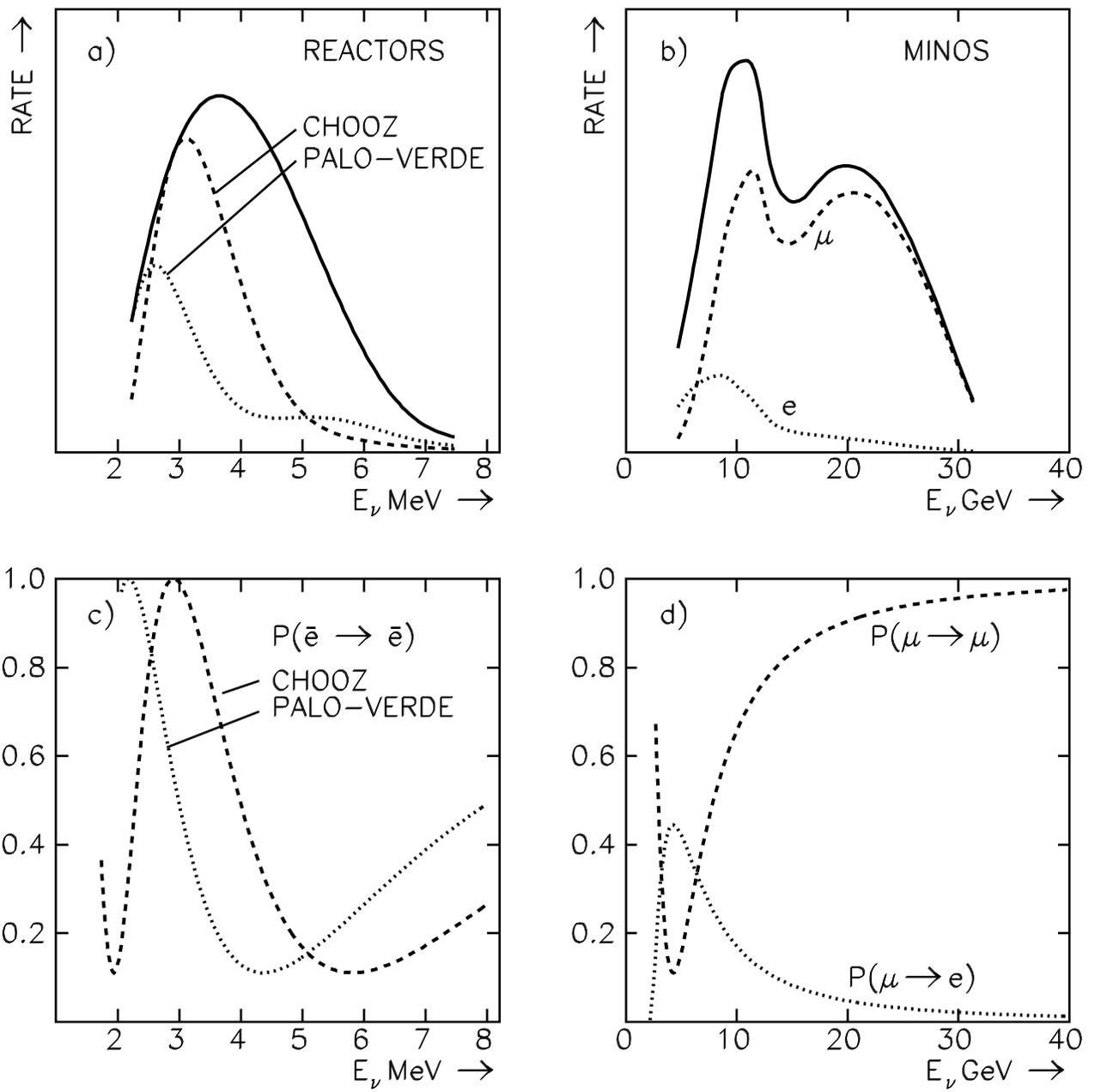
\noindent Figure~6.
The predicted event rate spectra 
in the threefold maximal mixing scenario \cite{HPS}
for a) The CHOOZ \cite{CHOOZ} and PALO-VERDE \cite{PALOV}
reactor experiments and b) the MINOS \cite{MINOS} 
long-baseline accelerator experiment.
Corresponding predictions c) and d) for
disappearance and appearance probabilities are also shown,
as a function of neutrino energy.

\end{document}